# Candidate Identification and Interference Removal in SETI@home

*Eric J. Korpela, Jeff Cobb, Matt Lebofsky, Andrew Siemion,
Joshua Von Korff, Robert C. Bankay, Dan Werthimer,
David Anderson*

## 1.0. Introduction

SETI@home, a search for signals from extraterrestrial intelligence, has been recording data at the Arecibo radio telescope since 1999. These data are sent via the Internet to the personal computers of volunteers who have donated their computers' idle time toward this search. The SETI@home client software, which runs on these computers, corrects the data for a wide variety of possible accelerations of the transmitter or receiver ranging from -100 Hz/s to 100 Hz/s. At each possible Doppler drift rate, the software performs a sensitive analysis to detect four types of potential signals: (1) narrowband continuous wave signals, (2) narrowband signals that match the Gaussian profile expected as an extraterrestrial signal drifts through the telescope field of view, (3) repeating pulses found using a fast folding algorithm, and (4) signals representing a series of three signals at constant frequency, evenly spaced in time (Korpela, Werthimer, Anderson, Cobb, and Lebofsky 2001).

To date, SETI@home volunteers have detected more than 4.2 billion potential signals. (http://setiathome.berkeley.edu/sci_status.html) While essentially all of these potential signals are due to random noise processes, radio frequency interference (RFI), or interference processes in the SETI@home instrumentation, it is possible that a true extraterrestrial transmission exists within this database. Herein we describe the process of interference removal being implemented in the SETI@home post-processing pipeline, as well as those methods being used to identify candidates worthy of further investigation.

## 2.0. Candidate Identification

Several properties make a candidate worthy of reobservation. Primarily, a good candidate should be persistent in its position in the sky. If we detect a frequency from a certain sky position, and detect an identical frequency from a point on the sky many degrees away, there are two possibilities: an extraterrestrial civilization has multiple beacons separated by hundreds of light years, all of which are Doppler corrected for the motions of the planet Earth, or we've detected a source of terrestrial interference. The latter is, of course, far more probable.

A good candidate should be persistent in time. For example the "Wow!" signal (Gray and Marvel 2001) had extremely high power, and it had the appropriate Gaussian profile for a point source drifting through the telescope's field of view, but despite repeated attempts at follow-up detections it has never been seen again. That makes it unlikely that the "Wow!" signal was a high duty cycle extraterrestrial beacon.

A good candidate should be persistent in frequency. When examined again it should appear at a similar frequency (but perhaps not identical due to uncorrected Doppler effects). Allowing too large a frequency difference makes it more likely that random noise events or unrelated interference could be considered to be part of a candidate.

The SETI@home candidate identification ranks groups of signals by their persistence in time, their spatial proximity, their dissimilarity to signals generated by random noise processes, their dissimilarity to known interference sources, and their proximity to interesting celestial objects (nearby or solar-type stars, known planetary systems, etc.) It assigns a score based upon the probability that the set of signals seen from a point in the sky would occur due to random noise processes, with lower scores being better.

Early in the project, candidate identification was an arduous process that was undertaken at intervals ranging from six months to more than a year. Because this process would access every signal in the SETI@home database several times, it was very I/O intensive and would require months to complete. To remove this shortcoming, we have designed a Near-Time Persistency Checker (NTPCkr).

The SETI@home pipeline keeps track of incoming potential signal locations by pixelating the sky in an equal area pixelization scheme. When a signal comes in, the corresponding sky pixel is marked as "hot" and given a time-stamp. Since a given area of sky tends to be observed several times in a short period, this pixel is allowed to "cool" for several weeks. At this point, if no further signals for that pixel are received, it is marked as ready for analysis.

The NTPCkr examines the signals within that pixel and adjacent pixels to determine a candidate score based upon the above criteria. It is our goal that the score represent the probability that the set of potential signals associated with the candidate could arise due to random noise processes. The existing candidates are ranked in order of this score from lowest (least noise-like) to highest (most noise-like).

## 3.0. Interference Removal

In the past, it has been our practice to perform interference removal on the entire set of potential signals detected by our instruments. Again, this method requires that the entire database be examined multiple times, which is inefficient.

Because narrowband correlations are very unlikely to occur due to random noise processes, candidate groups containing interference are ranked very highly on our candidate lists. Therefore, we now run interference rejection on candidate groups in order of their ranking. A candidate containing a lot of interference will have a good (low) score because it is not noise-like. The interference removal process will remove many of the non-noise-like signals, resulting in a candidate that is more noise-like, and thereby increasing (worsening) the score.

The interference removal techniques we use are independent and, because of the random access nature of the database, can be run in any order. After interference rejection, the candidate position is again marked as ready for analysis by the NTPCkr.

### 3.1. Radar Removal

By far, the most common source of interference in the SETI@home data set is radar stations on the island of Puerto Rico. Although these stations do not transmit within the 1.4GHz band received by the ALFA receiver used by SETI@home, signals from the radars do leak into the band, appearing as short duration, high intensity signals that change frequency rapidly with a large frequency component near the receiver central frequency. This component typically breaks up into multiple stable harmonics when seen in the recorded data. Fortunately, the radars are periodic, transmitting pulses of a few microseconds duration every few milliseconds. The pulse patterns and periods are known or can be measured. The Arecibo Observatory has build a radar blanking signal that is synchronized with the strongest radar and can be recorded with the data. However, this signal only removes the strongest radar and if the period or phase of that radar changes, it can take some time for the blanking signal to become resynchronized.

Therefore, we have built a software equivalent. This software radar blanker examines the data for radar pulses fitting the pattern of one of several known radars, determines the repetition period for that pattern and generates a signal indicating at what time the radar pulses should be present. Before distributing data to our volunteers, we replace these sections of data with a computer-generated noise-like signal. This typically results in a sensitivity loss of about 1.2 dB for strong narrowband signals with durations longer than the interpulse period. This loss is acceptable considering the alternative of filling the signal database with unwanted radar signals.

Our remaining interference mitigation methods are applied to the results returned by our volunteers after they have been inserted into our science database.

### 3.2. Zone Interference Removal

Zone Interference Removal removes signals that are contained within a "zone," which is a region of parameter space known to contain a large number of invalid signals. The parameters that define a zone can include a range of radio detection frequency, base-band frequency, period (for pulsed signals), detection time, the identity of the receiver, and the version of software used for various stages of the analysis process.

The top panel of Figure 3.1 shows the frequency distribution of 378,362,077 potential pulsed signals detected by SETI@home between July 5, 2006, and September 16, 2009. The vertical bands that are present indicate frequencies that are overrepresented and are probable RFI frequencies. We use a statistical analysis to determine which frequencies appear too frequently on differing sky positions to be due to noise processes. Those frequencies define the exclusion zones. Pulses determined to be within these zones (6.6 percent of the total) are shown in the middle panel. The lower panel shows the distribution of pulses that remain after those within zones have been removed.

The RFI frequency zones are typically quite narrow. We have identified 35,000 frequencies, covering less than 1 percent of our band which are subject to frequent interference. These zones contain between 5 and 20 percent of the detected signals depending upon signal type. As our software matures, our zone definitions are changing to better match interference characteristics. Signals determined to be within the zones are marked as interference and are excluded from future candidate scoring computations. This analysis can be done on other parameters (for example: pulse period or Doppler drift rate) to design RFI exclusion zones for those parameters as well.

### 3.3. Short-Term Fixed-Frequency Interference Removal

Some sources of interference are present at constant frequencies for periods of time ranging from hours to days, but not for sufficiently long to define a zone. Because celestial objects stay in our field of view for seconds to minutes, we can use this property to remove these sources of interference. By examining a time range around a potential signal we can calculate the probability of coincidence with another signal with similar frequency but seen at a different sky position. If this probability falls below a threshold ($\sim 10^{-4}$) we conclude that the signals are due to an interference source.

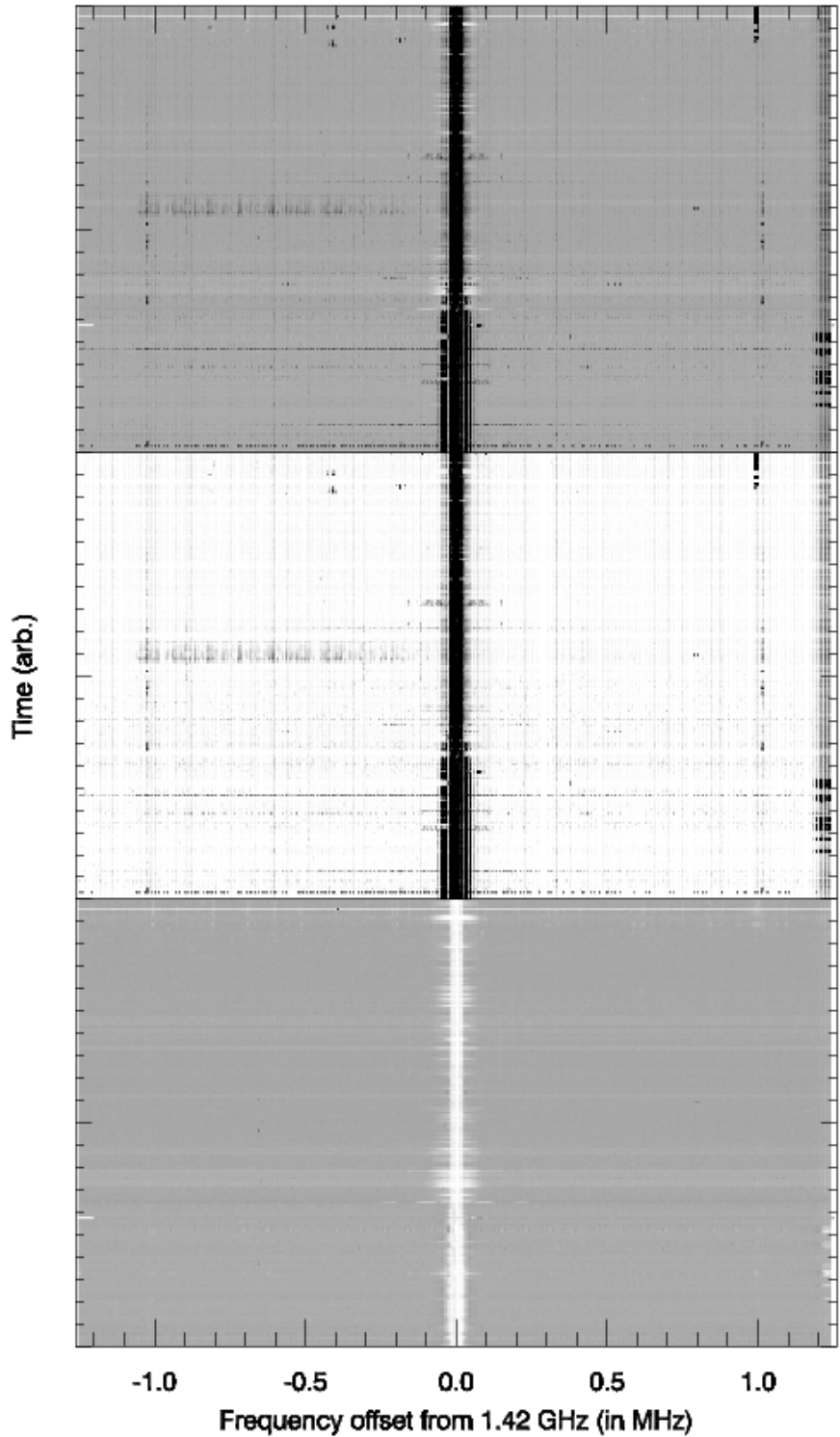

Figure 3.1: These plots show the frequency distribution of pulses detected by SETI@home. The upper panel shows all pulses. The middle panel shows pulses determined to be due to persistent interference sources. The lower panel shows the pulse frequency distribution after the interference has been removed. Note that some interference remains.

### 3.4. Removal of Interference that Drifts in Frequency

Some sources drift in frequency, even over short periods of time. For these methods we use the octant-excess drifting interference detection and removal method described by Cobb, Lebofsky, Werthimer, Bowyer, and Lampton (2000). Adjacent signals in time and frequency, but at different sky positions, are allocated into octants of frequency-time space surrounding the signal being examined. A significant statistical excess in an octant and the octant 180 degrees opposite indicates the presence of an RFI source drifting in frequency.

Again, a probability computation is used to determine the likelihood that this excess is due to random noise, and if this computation falls below a threshold, the signal being examined is marked as being due to interference.

### 3.5. Crowdsourced Interference Removal

The final stage of candidate identification requires examination of the top candidates by eye to detect forms of interference that might get past the first three layers of RFI removal. Because of the small amount of manpower available in the form of SETI@home staff members, we intend to develop a "crowdsourced" candidate investigation method. Similar to Stardust@home, it will use fabricated candidates, some containing RFI and others that are RFI clean, to train volunteers in identifying RFI and ranking candidates.

The lists of best candidates will be available online. Volunteers can then submit an opinion whether each of the signals making up the candidate is due to RFI. These votes will be used (in conjunction with the volunteer training scores) to modify the candidate score, which will alter the rankings.

## Acknowledgments

The SETI@home and Astropulse projects are funded by grants from NASA and the National Science Foundation, and by donations from the friends of SETI@home. Observations are made at the NAIC Arecibo Observatory, a facility of the NSF, administered by Cornell University.

## Works Cited

Cobb, J., M. Lebofsky, D. Werthimer, S. Bowyer, and M. Lampton. 2000. SERENDIP IV: Data acquisition, reduction, and analysis. In *Bioastronomy 99: A New Era in the Search for Life*, ASP Conference Series, 213: 485–89. ADS:2000ASPC..213..485C

Gray, R. H., and K. B. Marvel. 2001. A VLA search for the Ohio State "Wow." *The Astrophysical Journal*, 546: 1171–77. DOI:10.1086/318272

Korpela, E. J., D. Werthimer, D. Anderson, J. Cobb, and M. Lebofsky. 2000. SETI@home—Massively distributed computing for SETI, *Computing in Science and Engineering* 3 (1): 78–83. DOI:10.1109/5992.895191